\documentclass[preprint]{ametsocjmk}

\usepackage{graphicx}
\usepackage{amsmath}

\providecommand\bnabla{\boldsymbol{\nabla}}
\providecommand\bcdot{\boldsymbol{\cdot}}
\newcommand\dr{\mathrm{d}}
\newcommand\er{\mathrm{e}}
\newcommand\ir{\mathrm{i}}

\authoraddr{Fabrice Ardhuin, Centre Militaire d'Oc{\'e}anographie,\
Service Hydrographique et Oc\'{e}anographique de la Marine, 29609 Brest, France
\\E-mail: ardhuin@shom.fr}
\slugcomment{{\bf Fabrice's Draft:} \today}

\begin{document}

\title{Commentary on `The Three-Dimensional Current and Surface Wave Equations' \\ by George Mellor}

\author[1]{Fabrice Ardhuin}
\author[2]{Alastair D. Jenkins}
\author[3]{Konstadinos A. Belibassakis}
\affil[1]{Service Hydrographique et Oc\'{e}anographique de la
Marine, Brest, France} \affil[2]{Bjerknes Centre for Climate
Research, Geophysical Institute, Bergen, Norway}
\affil[3]{Technological Educational Institute of Athens, Greece}

\begin{abstract}
The lowest order sigma-transformed momentum equation given by
Mellor (J. Phys. Oceangr. 2003) takes into account a
phase-averaged wave forcing based on Airy wave theory.  This
equation is shown to be generally inconsistent due to inadequate
approximations of the wave motion. Indeed the evaluation of the
vertical flux of momentum requires an estimation of the pressure
$p$ and coordinate transformation function $s$ to first order in
parameters that define the large scale evolution of the wave
field, such as the bottom slope. Unfortunately there is no
analytical expression for $p$ and $s$ at that order. A numerical
correction method is thus proposed and verified. Alternative
coordinate transforms that allow a separation of wave and mean
flow momenta do not suffer from this inconsistency nor require a
numerical estimation of the wave forcing. Indeed, the problematic
vertical flux is part of the wave momentum flux, thus distinct
from the mean flow momentum flux, and not directly relevant to the
mean flow evolution.
\end{abstract}

\section{Introduction}
Wave-induced motions are of prime importance \vspace{0.25cm} in
the upper ocean, and in the coastal ocean (e.g. Ardhuin et al.
2005\nocite{Ardhuin&al.2005a} for a recent review). Therefore,
usual three-dimensional primitive equations must be modified to
account for waves. Among such modified equations, those based on
surface-following coordinates provide physically sound definitions
of velocities right up to the free surface, allowing a proper
representation of surface shears and mixing on a vertical scale
smaller than the wave height (i.e. a few meters). Any change of
coordinate adds some complexity in the derivation, but the final
equations can be relatively simple because part of the advective
fluxes are removed, and boundary conditions may be simplified. A
new set of such equations was recently derived by Mellor
(2003\nocite{Mellor2003}) using a change of the vertical
coordinate only, arguably the simplest possible. Mellor's (2003)
set of equations was originally derived for monochromatic waves,
but it is easily extended to random waves (e.g. Ardhuin et al.
2004\nocite{Ardhuin&al.2004b}, eq. 8). Unfortunately, we show here
that these equations, in the form given by Mellor, are not
consistent in the simple case of shoaling waves without energy
dissipation. A modification is proposed to solve the problem, but
it requires a numerical evaluation of the wave forcing terms. This
difficulty is due to the choice of averaging, and the same problem
arises with the alternative Generalized Lagrangian Mean equations
of Andrews and McIntyre (1978, eq. 8.7a, hereinafter aGLM). Both
Mellor's and aGLM equations describe the evolution of a momentum
quantity that contains the three-dimensional wave
(pseudo)-momentum (hereinafter called `wave momentum' for
simplicity, see McIntyre 1981\nocite{McIntyre1981} for details).
Writing an evolution equation for this quantity requires an
explicit description of the complex vertical fluxes of wave
momentum that are necessary to maintain the vertical structure of
the wave field in the surface gravity waveguide.

\section{The problem: wave motions and wave-following vertical coordinates}
We discuss here the simple case of monochromatic waves of
amplitude $a$ and wavenumber $k$ propagating in the horizontal $x$
direction, with all quantities uniform in the other horizontal $y$
direction. The surface and bottom elevations are $\eta(x,t)$ and
$-h(x)$, respectively, so that the local mean water depth is
$D(x,t)=h(x)+\overline{\eta(x,t)}$, with the overbar denoting an
Eulerian average over the wave phase. We shall assume that the
maximum surface slope is a small parameter $\varepsilon_1 = ka \ll
1$, and that the Eulerian mean current $\overline{u}$ in the
$x$-direction is uniform over the depth. Thus $\omega$ will denote
the radian wave frequency related to $k$ by the linear wave
dispersion relation (e.g. Mei 1989\nocite{Mei1989}),
\begin{equation}
\omega=k \overline{u} + \sigma = k \overline{u} + \left[g k \tanh
(kD)\right]^{1/2}\label{disp}.
\end{equation}

Finally, we assume that the water depth, current and wave
amplitude change slowly along the $x$-axis with a slowness
measured by a second small parameter $\varepsilon_2$ taken to be
the maximum bottom slope. We thus assume $|(\partial D/\partial
x)| \leq \varepsilon_2$, $|(\partial a/\partial x)| \leq
\varepsilon_2$, $|(\partial \overline{u}/\partial x)/(\sigma)|
\leq \varepsilon_2$, $|(\partial a/\partial t)/(\sigma a)| \leq
\varepsilon_2$, $|k (\partial \overline{u}/\partial t)/\sigma^2|
\leq \varepsilon_2$, $|(\partial D/\partial t)k/\sigma| \leq
\varepsilon_2$. The conditions on the bottom slope and current
gradients are consistent with the condition on the wave amplitude
gradient because in steady conditions the wave amplitude would
change due to shoaling over the current and/or bottom.

The vertical coordinate $z$ is implicitly transformed into Mellor's $\varsigma$
coordinate through
\begin{equation}
z=s\left(x,\varsigma,t \right)=\overline{\eta}+\varsigma D + \widetilde{s}
\label{sMellor}
\end{equation}
with $\widetilde{s}$ defined by Mellor's eq. (23b)  as
\begin{equation}
\widetilde{s}= \widetilde{s}_0= a F_{SS} \cos\left(kx-\omega
t\right)\label{stilde}
\end{equation}
and the vertical profile function $F_{SS}$ defined by
\begin{equation}
F_{SS}=\frac{\sinh \left[kD\right(1+\varsigma \left) \right] }{\sinh \left(
kD\right) } =\frac{\sinh \left[k\right(z+h \left) \right] }{\sinh \left(
kD\right) } +O(\frac{a}{D}).
\end{equation}

The coordinate transformation from $z$ to $\varsigma$ has the very
nice property of following the vertical wave-induced motion, at
least for linear waves on a flat bottom, and to first order in
$\varepsilon_1$. In that case the iso-$\varsigma$ surfaces are
material surfaces, and the fluxes of horizontal momentum through
one of these surfaces are simply correlations of pressure $p$
times the slope of that surface $\partial s/\partial x$ (figure
1.c), which replaces the wave-induced advective flux $u w$ in an
Eulerian point of view (figure 1.a). More generally, when
averaging is performed following water particles over their
trajectory (Lagrangian) or over their vertical displacement
(Mellor-sigma), the corresponding advective flux of momentum $u_i
u_j$  is replaced by a modified pressure force (figure 1).
\footnote{For the Generalized Lagrangian Mean (GLM) only the
contributions to lowest order in $\varepsilon_1$ are indicated.
Indeed, in GLM the wave-induced advective flux is not strictly
zero, but of higher order, since the average only follows a
zero-mean displacement with a residual advection, contrary to a
truly Lagrangian mean with zero advection (e.g. Jenkins 1986).}
\begin{figure}[htb]
\label{avg}
 \vspace{9pt}
\centerline{\includegraphics[width=7.6cm]{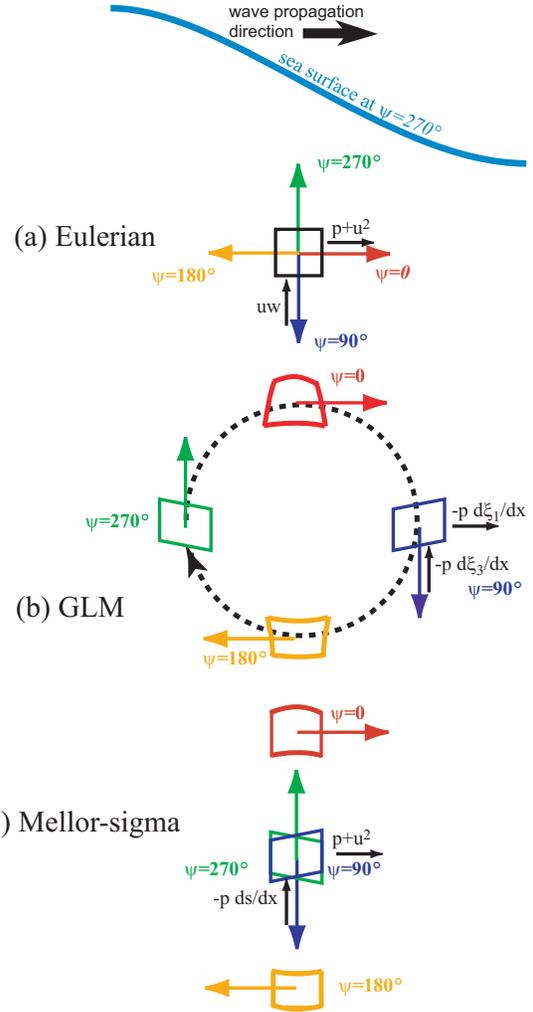}}
 \caption{Wave-induced fluxes of horizontal momentum in Eulerian, Generalized Lagrangian, and
 Mellor-sigma averages of the flow. Viscous or turbulent
 fluxes are neglected for simplicity. Distorted squares represent an elementary fluid volume and its position at four phases of the wave cycle,
 and the
 large arrow indicate the local wave orbital velocity.
 The horizontal and vertical fluxes of the horizontal momentum are represented by smaller
 arrows. Their expression are given to lowest order, without
 Jacobian corrections due to a change of volume (e.g. this results
 in $u^2+p$ becoming Mellor's $u^2+p \partial s/\partial
 \varsigma$).
}
\end{figure}

Using his coordinate transform, Mellor (2003) obtained a
phase-averaged equation for the drift current $U=\widehat{u}+u_S$
where $u_S$ is the Stokes drift, i.e. the mean velocity of water
particles induced by fast wave-induced motions. $U$ is strictly
defined as the phase-average particle drift velocity when
following the up-and-down wave motion, and $\widehat{u}=U-u_S$ is
a quasi-Eulerian mean current (Jenkins 1986, 1987). Below the wave
crests $\widehat{u}$ is equal, to second order in the wave slope,
with the Eulerian mean current $\overline{u}$ (figure 2).

\begin{figure}[htb]
 \vspace{9pt}
\centerline{\includegraphics[width=7.6cm]{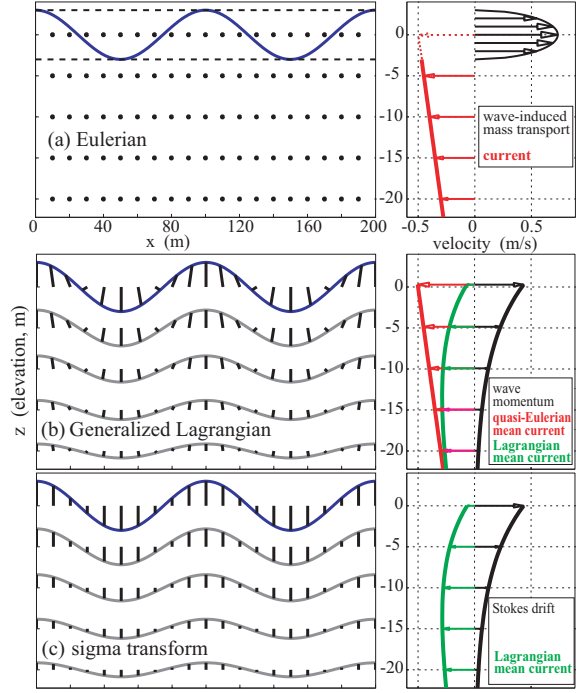}}
 \caption{Averaging procedures (left) and examples of resulting velocity profiles (right) in the
 case of (a) Eulerian averages,
 (b) the Generalized Lagrangian Mean, (c) sigma transforms (Mellor 2003).
 The thick black bars connect the fixed points $(x,z)$ where the average field is evaluated, to the displaced
 points $(x,z)+(\xi_1,\xi_3)$ where the instantaneous field is evaluated. For averages in moving coordinates the
 points $(x,z)+(\xi_1,\xi_3)$  at a given vertical level $\xi$ are along the gray lines. The drift velocity is the sum
 of the (quasi-Eulerian) current and the wave-induced mass transport.
 In the present illustration an Airy wave
 of amplitude 3~m and wavelength 100~m in 30~m depth, is superimposed on a hypothetical current of velocity
 $u(z)=-0.5-0.01 z$~m/s for
 all $z< \zeta(x)$. The
 quasi-Eulerian current profile is not represented in (c) since it is not directly given in Mellor's theory, although it can obviously be obtained by taking the
 difference of the other two profiles.}
 \label{figavg1}
\end{figure}

Mellor's horizontal mean momentum equation (34a) is reproduced
here for completeness, in our conditions with a flow restricted to
the vertical $x,z$ plane, a constant water density, no Coriolis
force, and no turbulent fluxes and the atmospheric mean pressure
set to zero (wind-wave generation due to air pressure fluctuations
is absorbed in $F_{x3}$),
\begin{equation}
\frac{\partial DU}{\partial t}+\frac{\partial DU^2}{\partial
x}+\frac{\partial \Omega U}{\partial \varsigma}+g D\frac{\partial
\widehat{\eta}}{\partial x}= F_{xx}+F_{x3}.\label{MellorMom}
\end{equation}

On the right-hand the first other term
\begin{equation} F_{xx}=-\frac{\partial S_{xx}}{\partial
x}=- \frac{\partial }{\partial x}\left(\overline{D
\widetilde{u}^2+\widetilde{p} \frac{\partial
\widetilde{s}}{\partial \varsigma}}\right)
\end{equation}
 represents the convergence of a horizontal flux of horizontal momentum that
accelerates the mean drift velocity $U$.

The other term
\begin{equation} F_{x3}=-\frac{\partial
S_{x3}}{\partial \varsigma}=\frac{\partial}{\partial \varsigma}
\left(\overline{\widetilde{p}
\partial \widetilde{s}/\partial x} \right)
\end{equation}
 represents a similar convergence of a vertical flux of horizontal momentum.

Defining $g$ as the acceleration due to the apparent gravity,
$\widetilde{p}$ and $\widetilde{s}$ are of the order of $g a$ and
$a$ respectively.  In general $\widetilde{p}$ and $\widetilde{s}$
are almost in phase, thus the flux $S_{x3}$ is of the order of $g
a \, \partial a/\partial x$, and the force $F_{x3}$ is of the
order of $g D k a \, \partial a/\partial x$. Thus, in the case of
shoaling waves, $F_{x3}$ is of the order of $g D \varepsilon_1^2
\varepsilon_2$.

Mellor estimated the vertical momentum flux $S_{x3}$ from
(\ref{stilde}) and the corresponding lowest order wave-induced
kinematic pressure on $\varsigma$ levels\footnote{This pressure
includes a hydrostatic correction due to the vertical
displacement.},
\begin{equation}
\widetilde{p}=\widetilde{p}_0= g a \left(F_{CC}-F_{SS}\right) \cos(kx - \omega
t) \label{p},
\end{equation}
$g$ is the acceleration due to the apparent gravity, and the vertical profile
function $F_{CC}$ is defined by
\begin{equation}
F_{CC}=\frac{\cosh \left[kD\right(1+\varsigma \left) \right] }{\cosh \left(
kD\right) }=\frac{\cosh \left[k\right(z+h \left) \right] }{\cosh \left(
kD\right) } +O(\frac{a}{D}).
\end{equation}

For non-dissipating shoaling waves, the right hand side terms of
eq. (\ref{MellorMom}) are of order $g D \varepsilon_1^2
\varepsilon_2$. The estimation of $F_{x3}$ thus requires the
knowledge of $\widetilde{p}/(gD)$ and $k \widetilde{s}$ to order
$\varepsilon_1 \varepsilon_2$, for which Airy theory is
insufficient. In particular, this estimation demands a formal
definition of $\widetilde{s}$, not given by Mellor (2003).
Further, eq. (7) is only valid if the wave-induced velocity
$\widetilde{\Omega}$ through $\varsigma$ levels is zero, or at
least, yields a negligible flux $\overline{\widetilde{\Omega}
\widetilde{u}}$ and a negligible mean Jacobian-weighted vertical
velocity $\Omega=\overline{\widetilde{\Omega}/(1+\partial
s/\partial \varsigma/D)}$. This is not the case over a sloping
bottom with Mellor's (2003) $s$ function.

\subsection{Formal definition of the the coordinate change $\widetilde{s}$}
For a general surface $\varsigma$ defined implicitly by
$z=s(x,\varsigma,t)$, the $\varsigma$ velocity component
$\widetilde{\Omega}$ is
 (e.g. Mellor 2003 eq. 20),
\begin{eqnarray}
\widetilde{\Omega}&=&\frac{\dr (z-s)}{\dr t} - \overline{\frac{\dr
(z-s)}{\dr t}}
\nonumber\\
&=&\widetilde{w}-\widetilde{u} \frac{\partial
\overline{s}}{\partial x}-\widehat{u} \frac{\partial
\widetilde{s}}{\partial x}-\frac{\partial \widetilde{s}}{\partial
t},
\end{eqnarray}
 with $\overline{s}=\overline{\eta}+\varsigma D$.

 In the spirit of Mellor's  (2003) derivation, the $\varsigma$ levels should be material surfaces for
 wave-only motions, so that one may neglect the vertical flux of momentum $\overline{\left(U+\widetilde{u}\right) \widetilde{\Omega}/(1+\partial s/\partial
\varsigma/D)}$.

Using the wave-induced vertical and horizontal displacements,
$\xi_3(x,\varsigma,t)$ and $\xi_1 (x,\varsigma,t)$, defined by
$\partial \xi_i/\partial t=\widetilde{u}_i (x_1+\xi_1,z+\xi_3,t)$,
we redefine the wave part of $s$,
\begin{equation} \widetilde{s}'\equiv \xi_3 -\xi_1 \frac{\partial \overline{s}
}{\partial x}.\label{sprime}
\end{equation}
The first term $\xi_3$ corresponds to Mellor's definition while
the second is a $O(\varepsilon_2)$ relative correction. This
definition yields a wave-induced vertical velocity
$\widetilde{\Omega}=-\widehat{u} \partial \widetilde{s}'/{\partial
x}$ through the iso-$\varsigma$ surfaces redefined by
 $z=s\left(x,\varsigma,t \right)=\overline{\eta}+\varsigma D +
 \widetilde{s}'$. If $\widehat{u} \ll \widetilde{u}$, as in the examples below, then $\widetilde{\Omega}$ is of a higher order
 compared to that given by Mellor's  (2003) $\widetilde{s}$ (eq. \ref{sMellor}).

\subsection{Wave-induced vertical displacements and pressure over a sloping bottom}
A WKBJ approximation using Airy's theory is sufficient for
estimating $\partial F_{xx}/\partial x$ because the horizontal
gradient of any wave-averaged quantity $\phi$ is of order  $
\varepsilon_2 k \phi$. On the contrary, the other force $F_{x3}$
is affected by modifications $\widetilde{s}'_1$ and
$\widetilde{p}_1$ to the local-flat-bottom solutions
$\widetilde{s}'$, and $\widetilde{p}$.

For small bottom slopes, $\widetilde{s}'_1$ and $\widetilde{p}_1$
are expected to be of the order of $ \varepsilon_2 \widetilde{s}'$
and $ \varepsilon_2\widetilde{p}$, i.e. of order $a \varepsilon_2$
and $g a \varepsilon_1 \varepsilon_2$, respectively. Thus
$\partial \widetilde{s}'_1/\partial x$ is of order $k a
\varepsilon_2=\varepsilon_1 \varepsilon_2$, and is expected to be
in phase with the wave-induced pressure
 (\ref{p}), of order $g a$, giving another term of
 order $g D \varepsilon_1^2 \varepsilon_2$ omitted by Mellor in his estimation of $\partial
(\overline{\widetilde{p}
\partial \widetilde{s}/\partial x})/\partial \varsigma$. The
modification of the pressure can be obtained from the modification
of the velocity potential, and it may be in phase with $\partial
\widetilde{s}_0/\partial x$, thus also contributing at the same
order to $F_{x3}$.

In order to be convinced of the problem, one may consider the case
of steady monochromatic shoaling waves over a slope without bottom
friction, viscosity or any kind of surface stress. We also neglect
the Coriolis force. In this mathematical experiment, the flow is
purely irrotational. We consider that the non-dimensional depth
$kH$ is of order 1, and that there is no net mass flux across any
vertical section. In that case the mean current and the Stokes
drift are of the same order, i.e. of the order $C \varepsilon_1^2$
with $C$ the phase speed. The mean current exactly compensates the
divergence of the wave-induced mass transport, and the mean sea
level is lower in the area where the wave height is increased
(Longuet-Higgins 1967\nocite{Longuet-Higgins1967})
\begin{equation}
\widehat{\eta}(x)= - \frac{k E}{\sinh (2 k D)}+  \frac{k_0 E_0}{\sinh (2 k_0
D_0)}
\end{equation}
where the $0$ subscript correspond to quantities evaluated at the
offshore boundary of the domain.

Since wave forcing is steady, the Eulerian mean current response
is steady (e.g. Rivero and Sanchez-Arcilla 1994, McWilliams et al.
2004, Lane et al.
2006\nocite{Rivero&Arcilla1995,McWilliams&al.2004,Lane&al.2007}),
 and thus the Lagrangian mean current is also steady. Thus the first term in
(\ref{MellorMom}) is zero and the second is of order $ D C^2
\varepsilon_1^4 \varepsilon_2 /D \simeq g D \varepsilon_1^4
\varepsilon_2$. The vertical mean velocity $\Omega$ can be
estimated from the steady mass conservation equation,
\begin{equation}
\frac{\partial D U}{\partial x}+ \frac{\partial \Omega}{\partial \varsigma} =0
\end{equation}
where the first term is of order $D C \varepsilon_1^2
\varepsilon_2/D$ and the second is of order $\Omega$. Thus the
third term in (\ref{MellorMom}) is of order  $C^2 \varepsilon_1^4
\varepsilon_2 \simeq g D \varepsilon_1^4 \varepsilon_2$. The
remaining terms in (\ref{MellorMom}) are of order $\varepsilon_1^2
\varepsilon_2$, giving the lowest order momentum balance
\begin{equation} \stackrel{\mathrm{Feta}}{-D \frac{\partial }{\partial x}\left(
g {\overline{\eta}}\right)}\stackrel{\mathrm{Fxx}}{-\frac{\partial S_{x x}
}{\partial x}}+\stackrel{\mathrm{Fx3}}{\frac{\partial }{\partial \varsigma}
\overline{\widetilde{p}
\partial \widetilde{s}/\partial x}}=0.\label{balance}
\end{equation}
For reference the corresponding lowest order Eulerian mean balance
is (e.g. Rivero and Sanchez-Arcilla 1994, Lane et al.
2006\nocite{Rivero&Arcilla1995,Lane&al.2007})
\begin{equation}
- \frac{\partial }{\partial x}\left( g
{\overline{\eta}}-\overline{\widetilde{w}^2}\right)-\frac{\partial
\overline{\widetilde{u}^2} }{\partial x}-\frac{\partial
\overline{\widetilde{u} \widetilde{w}} }{\partial
z}=0,\label{Eul_balance}.
\end{equation}
Only the hydrostatic pressure gradient is present in both the
Eulerian and Mellor-sigma balances, because the other terms
represent a different balance, including wave momentum in the
latter (see figure 2).

Equation (\ref{balance}) is now tested numerically. We take a
Roseau-type bottom profile (1976\nocite{Roseau1976}) defined by
$x$ and $z$ coordinates given by the real and imaginary part of
the complex function
\begin{equation} Z(x')=\frac{h_1 (x'-\ir \alpha)+(h_2-h_1) \ln (1+\er^{x'-\ir
\alpha})}{\alpha}.\label{Roseau}\end{equation}

With $\alpha=15 \pi/180$, $h_1=6$~m and $h_2=4$~m (figure 1), and
a radian frequency $\omega=1.2$~rad~s$^{-1}$ (i.e. a frequency
$f=0.2$~Hz), the non-dimensional water depth varies between
$0.85<kH<1.1$. The reflection coefficient for the wave amplitude
is $1.4\times 10^{-9}$ (Roseau 1976\nocite{Roseau1976}), so that
reflected waves may be neglected in the momentum balance. We
illustrate the force balance obtained for waves with an offshore
amplitude $a_0=0.12$~m, which corresponds to a maximum steepness
$\varepsilon_1=ka=2.6\times 10^{-2}$ equal to the maximum bottom
slope $\varepsilon_2=\varepsilon_1$. The change in wave amplitude
is given by the conservation of the wave energy flux (see Ardhuin
2006\nocite{Ardhuin2006b} for a thorough discussion), and the wave
phase $\psi$ is taken as the integral over $x$ of the local
wavenumber, so that $\partial \psi/\partial x = k$. The various
terms are then estimated using second order finite differences on
a regular grid in $\varsigma$ coordinates, with 201 by 401 points
covering the domain shown in figure 3.a. The three terms in eq.
(\ref{balance}) are shown in figure 3.
\begin{figure}[htb]
\label{Mellor_plot}
 \vspace{9pt}
\centerline{\includegraphics[width=8cm]{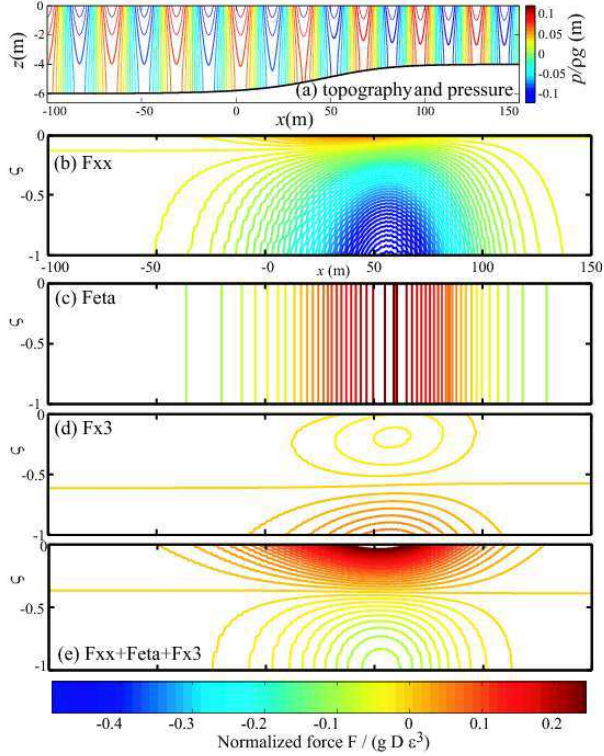}}
 \caption{(a) Snapshot of the pressure field for a slowly varying Airy wave over a the bottom topography
given by eq. (\ref{Roseau}). The forces in the balance (\ref{balance}) are
shown in panels b, c and d, with their sum in panel e, all estimated from
Mellor's analytical expressions. All forces have been
 normalized by $g D \varepsilon^3$. N.B. in the case shown here $\varepsilon_1=\varepsilon_2=\varepsilon$. }
\end{figure}

We have verified that the depth-integrated forces are in balance,
within 0.1\% of $F_{\mathrm eta}$. However, at most water depths
there is a large imbalance, of the order of the individual forces
( i.e. $g D \varepsilon_1^2 \varepsilon_2$) ,up to 180\% of
$F_{\mathrm eta}$. This contradicts the known steady balance
obtained from the Eulerian-mean analysis of Rivero and
Sanchez-Arcilla (1994).

For the case of shoaling waves without breaking the
three-dimensional equations of motion of Mellor (2003) are not
consistent to their dominant order, because of an improper
approximation of $S_{x3}$. This conclusion holds for any relative
magnitude of the wave and bottom slopes $\varepsilon_1$ and
$\varepsilon_2$.

\subsection{Wind-forced waves}
Clearly, any deviation of the wave-induced fields $s'$, $p$, and
$u$ from Airy-wave theory may have strong effects on the vertical
momentum flux term $S_{x3}$. Another example of such a situation,
correctly described by Mellor, is the case of wind-wave
generation. We briefly address it here because the full solution
can has not been given previously. Mellor focused on the wind-wave
generation contribution to the vertical momentum flux term
$\widetilde{p}
\partial \widetilde{s}/\partial x$ term. This equals the wave-supported wind stress at the sea surface, and, below, it explains the
growth of the wave momentum profile with the same profile as that
of the Stokes drift (Mellor 2003).

In horizontally uniform conditions, the wave amplitude is a
function of time only, and for the sake of simplicity we shall
solve the problem in the frame of reference moving at the velocity
at which the wave phase is advected by the current. We write the
wave-induced non-hydrostatic kinematic Eulerian pressure in the
form $\widetilde{p}_E=\widetilde{p}_{E0}+\widetilde{p}_{Ew}$, the
elevation as $\eta=\zeta_0+\zeta_w$
 and the velocity potential as $\phi=\phi_0+\phi_w$, in which the 0 subscript refers to the
primary waves, and the $w$ subscript refers to the added
components in the presence of wind forcing. Taking a primary
surface elevation of the form $\zeta_0= a \cos \psi$ with the the
phase $\psi=k x - \sigma t$, Mellor considered an atmospheric
kinematic pressure fluctuation in quadrature with the primary
waves
\begin{equation}
\widetilde{p}_a=-g \beta \frac{\rho_w}{\rho_a} a \sin \psi,\label{windforcing}
\end{equation}
with $\beta$ a small non-dimensional wave growth factor, and $\rho_w$ and
$\rho_a$ the densities of water and air respectively. He then assumed that the
water-side wave-induced pressure was of the form
\begin{equation}
\widetilde{p}_{\mathrm{Mellor}}= -g \beta a\frac{\cosh \left[k\right(z+h \left)
\right] }{\cosh \left( kD\right) } \sin \psi.\label{pwMellor}
\end{equation}

Implicitly $\widetilde{s}'_w$ is zero, and for his purpose $\phi_w$ was
irrelevant. We shall now also determine $\phi_w$. The continuity of dynamic
pressures at the surface is\footnote{Here the pressure is Eulerian. For
correspondance to Mellor's pressures on $\varsigma$ levels, one should take
$\widetilde{p}=\widetilde{p}_{E}-g \widetilde{s}'$.}
\begin{equation}
\widetilde{p}_{Ew}+g \widetilde{s}'_{w}= -g \beta a \sin \psi
\quad{\mathrm{at}}\quad z=\zeta.\label{surfpres}
\end{equation}

A solution is obtained by solving Laplace's equation with proper
boundary conditions,  to first order in $\beta$. The boundary
conditions include the Bernoulli equation,
\begin{equation}
    \frac{\partial \phi}{\partial t}=
   -g\zeta - \frac{\rho_a}{\rho_w} p_a,\quad \mbox{at\ }\quad z=\zeta,
    \label{Bernoulli vent}
\end{equation}
in which non-linear terms have been neglected because they are the sum of
products of the form $\bnabla \phi_0 \bcdot  \bnabla \zeta_0$, unchanged from
the case without wind, and terms of the form $\bnabla \phi_w \bcdot  \bnabla
\zeta_0$, which are negligible compared to the left-hand side terms for primary
waves of small slope. Similarly, the surface kinematic boundary condition is
linearized as
\begin{equation}
    \frac{\partial \phi }{\partial z}
    =\frac{\partial \zeta}{\partial t}  \quad \mbox{at\ }\quad
    z=\overline{\zeta}.\label{surfkin}
\end{equation}

The combination of both yields
\begin{equation}
    \frac{\partial ^{2}\phi}{\partial t^{2}}
    +g\frac{\partial \phi}{\partial z}
    =-\frac{\rho_a}{\rho_w}\frac{\partial \widetilde{p}_a}{\partial t}  \quad \mbox{at\ }\quad
    z=\overline{\zeta}.
    \label{surface vent}
\end{equation}
$\phi_w$ is also a solution of Laplace's equation with the bottom boundary
condition $\partial \phi_w / \partial z=0$ at $z=-h$. With the fully resonant
atmospheric pressure (\ref{windforcing}) envisaged by Mellor, one has
\begin{eqnarray}
    \zeta_0&=& a(t) \cos \psi, \\
    \phi_0&=& \frac{g a(t)}{\sigma} F_{CC} \sin \psi, \\
    {\widetilde p}_{E0}&=& g a(t) F_{CC} \cos \psi, \\
    \frac{{\mathrm d} a(t)}{{\mathrm d} t}&=&\frac{\beta \sigma a(t)}{2},\\
     \phi_w&=&\beta g \frac{a}{2 \sigma}  F_{CC} \left[A \cos
    \psi+B
    \cos \psi'\right] \label{phi_wind}
\end{eqnarray}
with $\psi'=\left(kx+\sigma t\right)$. The elevation and
under-water non-hydrostatic pressure corresponding to $\phi_w$ are
given by (\ref{surfkin}) and the linearized Bernoulli equation
\begin{eqnarray}
 \frac{\partial }{\partial t}\left( \phi_0 + \phi_w\right)
    = -  \widetilde{p}_{E0} - \widetilde{p}_{Ew}
\end{eqnarray}
yielding
\begin{eqnarray}
    \zeta_w&=&\beta \frac{a}{2} \left[\left(1-A\right) \sin \psi +B \sin \psi
    '\right] \\
    \widetilde{p}_{Ew}&=& g \beta \frac{a}{2} F_{CC} \left[-\left(1+A\right) \sin\psi+B \sin\psi' \right]. \nonumber\\\label{pw}
\end{eqnarray}
Mellor's expression  for $\widetilde{p}_w$  (eq. 18) is obtained
by replacing $\zeta_w$ and $\widetilde{p}_{Ew}$ in
(\ref{surfpres}), giving $A=1$. One may take $B=1$ to have
$\phi_w=0$ at $t=0$, or more simply $B=0$, which gives
$\zeta_w=0$, and $p_{Ew}=F_{CC} \widetilde{p}_a \rho_a/\rho_w$.
The choice of $B$ has no dynamical effect. In the present case
$\phi_w$ should give a contribution to $S_{x x}$ because it is in
phase with $\phi_0$, but this is a relative correction of order
$\beta$, thus negligible. To the contrary, the contribution of
$\widetilde{p}_{w}$ to $(\overline{\widetilde{p}
\partial \widetilde{s}/\partial x})$ is quite important, because for
uniform horizontal conditions this flux is otherwise zero.

\section{A solution to the problem ?}
Contrary to that particular wind-forcing term, there is no simple
asymptotically analytical correction for $\widetilde{p}$ and
$\widetilde{s}'$ that can account for the bottom slope and wave
field gradient. A major problem in this situation is that the wave
velocity potential becomes a non-local function of the water
depth. The velocity
 potential and pressure fields may only be investigated analytically for plane beds (e.g. Ehrenmark
 2005\nocite{Ehrenmark2005}) or specific bottom profiles.
 Numerical solutions for the three-dimensional wave
motion are generally found as infinite series of modes (e.g.
Massel 1993\nocite{Massel1993}). The velocity potential for any of
these modes satisfies Laplace's equation with a local vertical
profile $F_n$ proportional to $\cos(k_n z+k_n h)$ and a dispersion
relation $\sigma^2=g k_n \tan (k_n D)$. The local amplitudes of
these modes  are non-local functions of the water depth, and may
be obtained numerically with a coupled-mode model (Massel 1993).
This non-local dependance of the wave amplitude on the water depth
arises from the elliptic nature of Laplace's equation, satisfied
by the velocity potential in irrotational conditions. The series
of modes can be made to converge faster by adding a `sloping
bottom mode' that often accounts for a large part of the
correction and is a local function of the depth and bottom slope.
It is thus of interest to see if that correction only, without the
infinite series, may provide a first order analytical correction
to Mellor's momentum flux $S_{x3}$.

Following Athanassoulis and Belibassakis
(1999\nocite{Athanassoulis&Belibassakis1999}), one may define the velocity
potential for that mode as
\begin{eqnarray}
 \phi_{1}&=& - \frac{\dr h}{\dr x}a \sigma D F(z)  \cos \psi,
\end{eqnarray}
In order to satisfy the bottom boundary condition $w=\partial \phi_{1}/\partial
z= - \frac{\dr h}{\dr x} \partial \phi_{0}/\partial x$, the function $F$ should
verify $D {\dr F}(-h)/{\dr z}=1/\sinh(kH)$ and the satisfaction of the surface
boundary condition may be obtained with $F(0)={\dr F}(0)/{\dr z}=0$.
Athanassoulis and Belibassakis (1999) have used
\begin{equation}
    F=F_{\mathrm{AB}}\equiv\frac{1}{\sinh(kD)}\left[\left(\frac{{z}-\overline{\zeta}}{D}\right)^3+\left(\frac{{z}-\overline{\zeta}}
    {D}\right)^2\right],\label{FAB}
\end{equation}
and Chandrasekera and Cheung
(2001\nocite{Chandrasekera&Cheung2001}) have used
\begin{equation}
    F=F_{\mathrm{Ch}}\equiv\frac{1}{kD \sinh^2 (kD)}\left[1-\cosh(k{z}-k\overline{\zeta})\right].\label{FCh}
\end{equation}

With these choices $\phi_1$ does not satisfy exactly Laplace's equation, and
thus requires further corrections in the form of evanescent modes. An infinite
number of other choices is available, either satisfying Laplace's equation or
the surface boundary conditions, but never both, so that each of these
solutions is only approximate, and the exact solution is given by the infinite
series of modes, which can be computed numerically for any bottom topography
(e.g. Athanassoulis and Belibassakis 1999, Belibassakis et al. 2001, Magne et
al. 2006\nocite{Belibassakis&al.2001,Magne&al.2007}).

The vertical displacement and Eulerian pressure corrections are given by time
integration of the vertical velocity and the linearized Bernoulli equation,
\begin{eqnarray}
 \xi_{31}&=& \frac{\dr h}{\dr x}a  D \frac{\dr F}{\dr z}  \sin \psi \\
 \widetilde{p}_{E1}&=& \frac{\dr h}{\dr x}a  D F(z)  \sin \psi
\end{eqnarray}

Thus, in absence of wind forcing but taking into account the `sloping bottom
mode' to first order in the bottom slope, the wave-induced flux of momentum
through iso-$\varsigma$ surfaces is
\begin{eqnarray}
  \widetilde{p} \frac{\partial \widetilde{s}'}{\partial
x}=
\left(F_{CC}-F_{SS}\right)\left[\frac{ga}{2} \frac{\partial( a F_{SS})}{\partial x} \right. \nonumber\\
  + \left. \frac{g k a^2}{2 }\frac{\dr h}{\dr
 x} \left(D \frac{\dr F}{\dr z} + \varsigma F_{CS} \right)\right]\nonumber\\
 + \frac{g k a^2}{2} \frac{\dr h}{\dr
 x}\left[-F_{SC} F+F_{SS} \left(D \frac{\dr F}{\dr z} + \varsigma F_{CS} \right)\right],
 \nonumber\\
\end{eqnarray}
with $F_{CS}={\cosh \left[kD\right(1+\varsigma \left) \right] }/{\sinh \left(
kD\right) }$. The first line is the term given by Mellor (2003). The second
line arises from the correction due to the difference between $\widetilde{s}'$
and $\widetilde{s}$, and the third line arises due to corrections
$\widetilde{p}_1=\widetilde{p}_{E1}-gs'_1$ to the pressure on $\varsigma$
levels. These additional term are of the same order as the first term, and have
no flux at the bottom and surface. Thus the depth-integrated equations
including that term also comply with known depth-integrated equations (e.g.
Smith 2006\nocite{Smith2006a}).

In the case chosen here $F_{\mathrm{Ch}}$ gives a net momentum
balance closer to zero than  Mellor's (2003) original expression
(figure 4). However, the remaining error is significant. Thus one
cannot use only that mode, and the contribution of the evanescent
modes have to be computed, which can only be done numerically.
\begin{figure}[htb]
\label{test1}
 \vspace{9pt}
\centerline{\includegraphics[width=8cm]{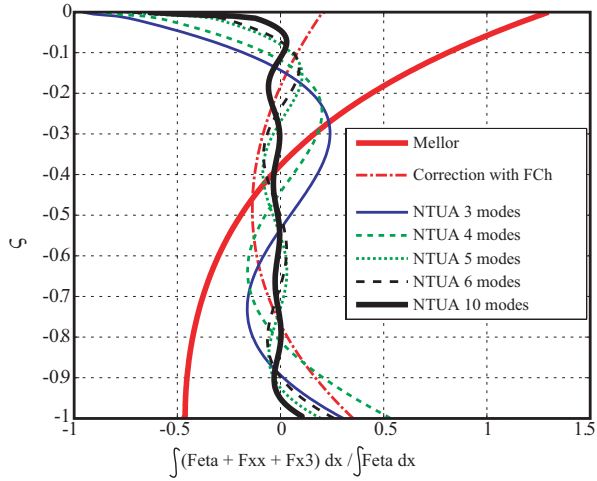}}
 \caption{Net forces in the momentum balance (\ref{balance}) for steady shoaling waves over a smooth bottom profile.
 The net force has been integrated over $x$ and normalized by a similar integration of the the hydrostatic pressure force
 $F_{eta}$. Several solutions are obtained. One corresponds to Mellor's orginal expression, one possible analytical correction using
 $F_{\mathrm{Ch}}$, and
 numerical estimations using the NTUA model, with various numbers of modes.}
\end{figure}

 A numerical evaluation
of the forces was performed using the NTUA model (Athanassoulis
and Belibassakis 1999). The NTUA solution was obtained in a domain
with 401 points in the horizontal dimension. For the small bottom
slope used here, the model contains a numerical reflection
$R=0.002$ much larger than the analytical value given by Roseau
(1976). However, this only introduces a modulation, in the $x$
direction, of the estimated forces. This modulation is significant
but still relatively smaller than the average. The net force
estimated from NTUA results is found to converge to the expected
force balance described by eq. (\ref{balance}) as the number of
evanescent modes is increased (figure 4). In this calculation the
values of $F_{xx}$ do not differ significantly from those
estimated using Mellor's analytical expressions, as expected. The
only significant difference between the NTUA numerical result with
10 modes, and Mellor's analytical expression is found in $F_{x3}$,
with a much stronger value near the surface in the numerical
result, allowing a balance with the strongly sheared $F_{xx}$
(figure 4).

\section{Conclusions}
Mellor (2003) changed  the vertical coordinate from $\varsigma$ to
$z$, using an implicit function $s$ in two parts,
$z=\overline{s}(x,y,\varsigma,t)+\widetilde{s}(x,y,\varsigma,t)$
with $\overline{s}$ changing only slowly in space and time and
$\widetilde{s}$ representing the faster wave-induced change of
vertical coordinate. If the $\varsigma$ levels are material
surfaces,  then the momentum flux $S_{x3}=\widetilde{p}^\xi
\partial \widetilde{s}/\partial x$ is the surface-following coordinate
counterpart of the Eulerian vertical momentum flux term
$\overline{\widetilde{u}\widetilde{w}}$ discussed by Rivero and
Arcilla (1995\nocite{Rivero&Arcilla1995}), with
$\widetilde{p}^\xi$ the wave-induced pressure at the displaced
position (in the surface-following coordinates). However,
$\widetilde{p}^\xi
\partial \widetilde{s}'/\partial x$ and $\overline{\widetilde{u}\widetilde{w}}$ do not represent the same physical quantity
since the former contains wave momentum, which is not included in
the latter.

Just like the Eulerian momentum flux $\overline{\widetilde{u}
\widetilde{w}}$ is modified by the bottom slope, wave amplitude
gradients, wind-wave generation, boundary layers, or vertical
current shears, these effects also modify $S_{x3}$. But in these
situations, the $\varsigma$ levels as defined by Mellor (2003) are
not material surfaces, and a missing Eulerian-like flux term
$\overline{\widetilde{\Omega} \widetilde{w}}$ would have to be
added to correct the momentum equations, with $\widetilde{\Omega}$
the wave-induced velocity across $\varsigma$ levels.
Alternatively, we propose to replace $\widetilde{s}$ with
$\widetilde{s}'$, defined by eq. (\ref{sprime}) such that
$\varsigma$ levels are closer to material surfaces, i.e. so that
$\widetilde{\Omega}$ is of a higher order.

Whether the original $\widetilde{s}$ or our corrected
$\widetilde{s}'$ is used, the wave-induced momentum flux $S_{x3}$
must be estimated to first order in the bottom slope
$\varepsilon_2$ for consistency. This requires an
$O(\varepsilon_2)$ estimation of both $\widetilde{p}^\xi$  and
$\widetilde{s}$ or $\widetilde{s}'$. Unfortunately there is no
analytical $O(\varepsilon_2)$ expression for the wave motion. Thus
Mellor's equations, even when corrected, require a
computer-intensive solution that is generally not feasible. For
example, Magne et al. (2006) only included a total of five modes
in their calculation of wave propagation over a submarine canyon.
In an example shown here, this small number of modes is
insufficient for an accurate estimation of wave-forcing terms.

The trouble with these equations can be avoided by using, instead,
equations of motion for the quasi-Eulerian velocity
$\widehat{u}=U-u_S$ (Jenkins 1986, 1987,
1989\nocite{Jenkins1987,Jenkins1986,Jenkins1989}). Such equations
have been obtained in the limit of vanishing wave amplitude using
an analytical continuation (e.g. using a Taylor expansion) of the
current profile across the surface (McWilliams et al.
2004\nocite{McWilliams&al.2004}). A general and explicit solution
can also be obtained from the exact Generalized Lagrangian Mean
(GLM) equations of Andrews and McIntyre
(1978a\nocite{Andrews&McIntyre1978a}) expanded to second order in
the surface slope $\varepsilon_1$ (Ardhuin et al., manuscript
submitted to Ocean Modelling, see
[http://arxiv.org/abs/physics/0702067]). In these, the equation
for the horizontal quasi-Eulerian momentum
 involves no flux term like $\overline{\widetilde{p}{\partial
\widetilde{s}}/{\partial x}}$ because this corresponds to the flux
$\overline{p^\xi
\partial \xi_3/\partial x}[1+O(\varepsilon)]$ of wave
momentum $u_S$ (Andrews and McIntyre
1978b{\nocite{Andrews&McIntyre1978b}, eq. 2.7b), not directly
relevant to the problem of mean flow evolution (see also Jenkins
and Ardhuin 2004\nocite{Jenkins&Ardhuin2004}). This flux of wave
momentum only appears in evolution equations for the total
momentum $U$, such as given by Mellor (2003), or the `alternative'
form of the GLM equations  (Andrews and McIntyre 1978, eq. 8.7a).

For that reason, the equations for the quasi-Eulerian velocity
$\widehat{u}$ are simple and consistent in their adiabatic form
(without wave dissipation), at least to lowest order in wave slope
and current vertical shear, for which analytical expressions exist
for the wave forcing terms. Further details on the relationships
between all these equations, and further validation against
numerical solutions of Laplace's equation can be found in Ardhuin
et al. (submitted manuscript).

{\it Acknowledgments.} This work was sparked by a remark of the
late and much missed Tony Elfouhaily. The open collaboration of
George Mellor and his criticisms of an early draft greatly
improved the present note, together with the feedback of anonymous
reviewers and many colleagues. Among them Dano Roelvink suggested
the evaluation of the torque resulting from Mellor's equations in
the vertical plane, and Jacco Groeneweg made a detailed critique
of early drafts.
 Initiated during the visit of A.D.J. at
SHOM, this research was supported by the Aurora Mobility Programme for Research
Collaboration between France and Norway, funded by the Research Council of
Norway (NFR) and The French Ministry of Foreign Affairs. A.~D.~J. was supported
by NFR Project 155923/700.


\bibliographystyle{ametsocjmk}
\bibliography{../references/wave}

\end{document}